**High-Performance Perovskite Photodetectors Based on $CH_3NH_3PbBr_3$ Quantum Dot/$TiO_2$ Heterojunction**

**Rajeev Ray, Nagaraju Nakka, and Suman Kalyan Pal[*]**

Advanced Materials Research Centre, India Institute of Technology Mandi, Kamand, Mandi-175075, Himachal Pradesh, India.

School of Basic Sciences, India Institute of Technology Mandi, Kamand, Mandi-175075, Himachal Pradesh, India.

AUTHOR INFORMATION

**Corresponding Author**

*E-mail: suman@iitmandi.ac.in; Phone: +91 1905 267040

**ORCID**

Suman Kalyan Pal: 0000-0003-2498-6217

**Abstract**

Organo-lead halide perovskite materials have opened up a great opportunity to develop high performance photodetectors because of their superior optoelectronic properties. The main issue with perovskite-only photodetector is severe carrier recombination. Integration of perovskite with high-conductive materials such as graphene or transition metal sulfides certainly improved the photoresponsivity. However, achieving high overall performance remains a challenge. Here, an improved photodetector is constructed by perovskite quantum dots (QDs) and atomic layer deposited (ALD) ultrathin $TiO_2$ films. The designed $CH_3NH_3PbBr_3$ QD/$TiO_2$ bilayer device displays inclusive performance with on/off ratio of $6.3\times10^2$, responsivity of 85 $AW^{-1}$, and rise/decay time of 0.09/0.11 s. Furthermore, we demonstrate that interface plays a crucial role in determining the device current and enhance the overall performance of heterostructure photodetectors through interface engineering. We believe that this work can provide a strategy to accelerate development of high-performance solution-processed perovskite photodetectors.

**1. Introduction**

Photodetectors are sensors having widespread applications in optical communications, imaging, and chemical/biological sensing [1]. At present, photodetectors made of inorganic semiconductors are ruling the markets [2]. The main disadvantages of these materials are non-flexibility, high processing cost and not environment–friendly, therefore, an alternative material could be a boon to the electronic industry [3]. Organic-inorganic halide perovskites are serious contenders for next-generation photovoltaic technology[4] owing to their extraordinary optoelectronic properties such as direct bandgap, high absorption coefficients, low non-radiative Auger recombination, small exciton binding energies and, long lifetime and high mobility of photocarriers[4c, 5]. Within a very short period (about ten years) the power conversion efficiency (PCE) of perovskite photovoltaic cells has reached 23.7% from 3.81%[6].

Recently, solution-processed perovskite-based photodetectors have shown great potential in photodetection[7]. Yang and co-workers reported a vertical perovskite photodetector[7d] that exhibits excellent light-detecting capability. However, lateral photodetectors are easy to fabricate because of quite a simple device structure; their intrinsic gain mechanism can lead to very high photosensitivity[8]. The single-layer (lateral) perovskite photodetectors show relatively high detectivity ($\approx 10^{12}$ Jones) and short response time (tens of milliseconds), but, low on/off ratio and poor electrical instability (induced by ion migration)[7b, 9]. The photocurrent in single-layer perovskite photodetector devices is relatively low because of fast recombination of the photogenerated carriers. These shortcomings are overcome in bilayer devices where perovskite, responsible for light absorption is placed on another material that transports charges. The carrier transport layer underneath the perovskite layer reduces recombination through efficient charge separation at the interface. Significant enhancement of photoresponsivity has been achieved in heterojunctions of perovskite and 2D materials such as graphene, $MoS_2$[10], and $WS_2$[11]. Graphene-based photodetectors[9c, 11a-d] offer very low on/off ratio, whereas detectors with $MoS_2$ possess low rise and decay times[11e, 11f].

Perovskite quantum dot (QD) based photodetectors are being quickly developed as well. Wang et al.[7a] fabricated bilayer photodetectors with all-inorganic perovskite QDs and mesoporous $TiO_2$. Epitaxially blended $CH_3NH_3PbBr_3$ ($MAPbBr_3$) QDs and ternary $PbS_xSe_{1-x}$ QDs have been used as active layer in photodetectors[12]. However, the performance of perovskite QD based photodetectors is limited by low photoresponsivity. In this article, we have fabricated MAPbBr$_3$QD/TiO$_2$ heterojunction on FTO substrate for photodetection. Mesoporous TiO$_2$ films have been prepared via atomic layer deposition (ALD) technique for better transport of photocarriers. The fabricated bilayer devices exhibit excellent performance in terms of photoresponsivity and response time. The performance of the photodetector is found to be highly dependent on the quality of the perovskite/TiO$_2$ interface. Moreover, UV-ozone treatment of the TiO$_2$ layer improves the detector performance.

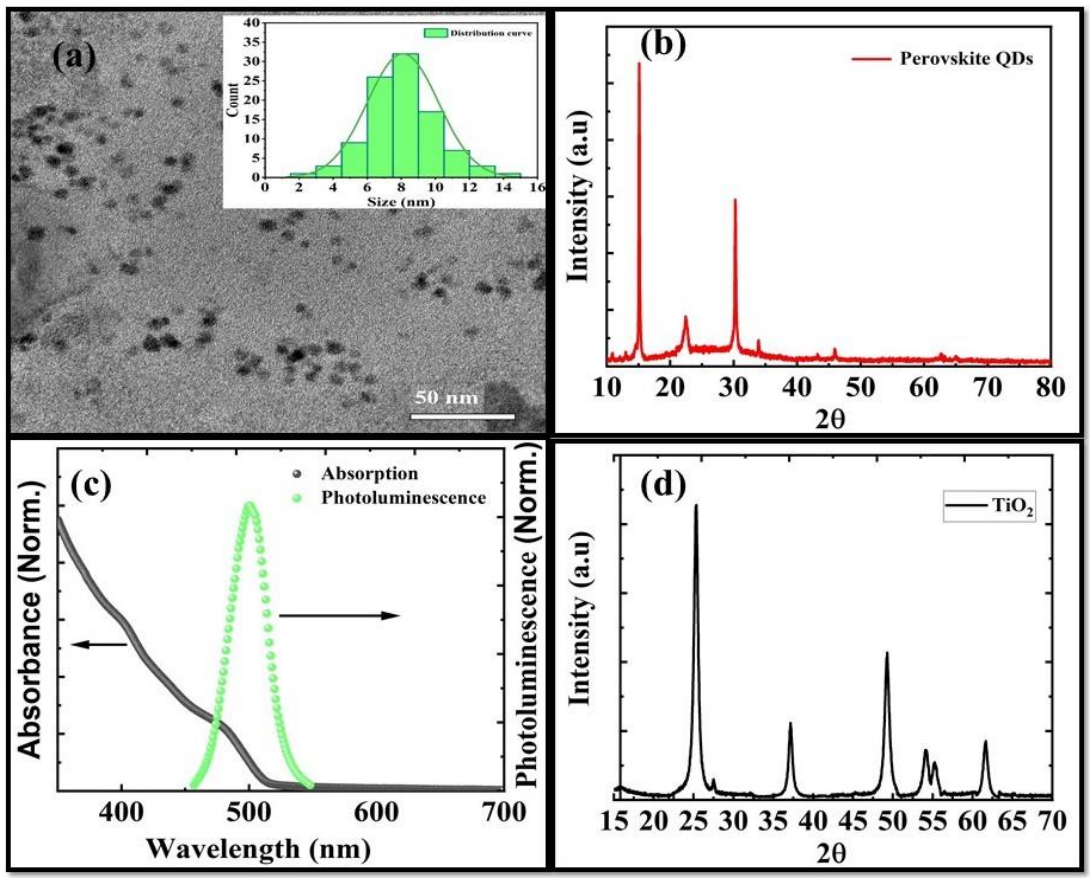

**Figure 1.** (a) TEM image of MAPbBr$_3$ QDs. The inset shows the particle size distribution of MAPbBr$_3$ QDs. (b) XRD spectrum of MAPbBr$_3$ QD film. (c) Absorption and PL spectra of MAPbBr$_3$ QD films. (d) XRD pattern of TiO$_2$ thin film prepared through ALD.

## 2. Results and Discussion

### 2.1. Characterization of MAPbBr$_3$ QDs and TiO$_2$ Thin Film

We prepared MAPbBr$_3$ QDs using facile solution process, which is given in the Experimental Section. The TEM image of perovskite QDs is illustrated in Figure 1a. It is seen from the inset of Figure 1a that the size of QDs is ~ 7.5 nm. Figure 1b depicts the X-ray diffraction (XRD) spectrum of the perovskite film. All the peaks (at 14.9°, 23.8°, 31.2°, and 46.04°) observed in the XRD spectrum are well matched to the tetragonal perovskite structure. No impurity peaks are observed other than those attributed to MAPbBr$_3$. The light absorption spectrum of MAPbBr$_3$ QD film is shown in Figure 1c. QDs exhibit strong absorption in the UV and visible light range with a peak at 490 nm. A strong PL is emitted from QDs at 520 nm (Figure 1c). The narrow FWHM of the PL band suggests uniform size distribution of QDs, which is also apparent from the TEM histogram (Figure 1a). The absorption edge of mesoporous TiO$_2$ was appeared at 388 nm (not shown) indicating bandgap energy of 3.2 eV. Figure 1d shows XRD pattern of ALD-TiO$_2$ film (annealed). The presence of 2θ peaks at 25.27° and 48.01° validates the anatase phase formation in TiO$_2$ films.

### 2.2. Single Layer vs Bilayer Devices

When MAPbBr$_3$ QD/TiO$_2$ heterojunction is used as the active material in photodetectors, the photoelectrical performance is significantly enhanced. Figure 2a presents the schematic device structure for MAPbBr$_3$ QD/mesoporous TiO$_2$ heterojunction photodetector on patterned FTO coated substrate. To demonstrate the superiority of heterojunction photodetectors, two devices: only TiO$_2$ (PDR) and MAPbBr$_3$ QD/TiO$_2$ (PD1) were fabricated. The thickness of the TiO$_2$ layer was 50 nm in both the devices. Figure 2b shows the symmetric waterfall type I-V characteristics of the photodetectors suggesting ohmic contact between TiO$_2$ film and electrodes, which is essential for good photoconductor and photodetector devices [13]. The dark current of the device PDR measured at 5 V is quite low; in the order of nanoampere. It is even lower in the heterojunction device PD1 because of the formation of a low-conductive depletion region at the perovskite /TiO$_2$ heterojunction. The photocurrent of PD1 increases dramatically up to 0.3 µA (at 5 V), which is more than one order of magnitude higher than that of PDR (Figure 2b). The low photocurrent in PDR can be explained from the fact that TiO$_2$ absorbs less light in the visible region due to very high bandgap as mentioned before. On the other hand, in the bilayer device, a large number of photocarriers is generated in the perovskite layer upon light illumination because of its strong absorption in the visible light range. As shown in figure 2c, MAPbBr$_3$/TiO$_2$ heterojunction possesses type II band alignment that causes photogenerated electrons to transfer into the underlying TiO$_2$ film by spatially separating them from photogenerated holes. The separation of photogenerated carriers via electron transfer can reduce carrier recombination leading to low PL from perovskite QDs. It is proved strongly from the PL quenching of perovskite QDs (Figure 2d) that the recombination of electron-hole pairs in QD film is suppressed due to the electron transfer to TiO$_2$ layer. The injection of electron into TiO$_2$ can enhance the conductivity of the TiO$_2$ layer, which is the key requirement behind high-performance of heterojunction photodetectors.

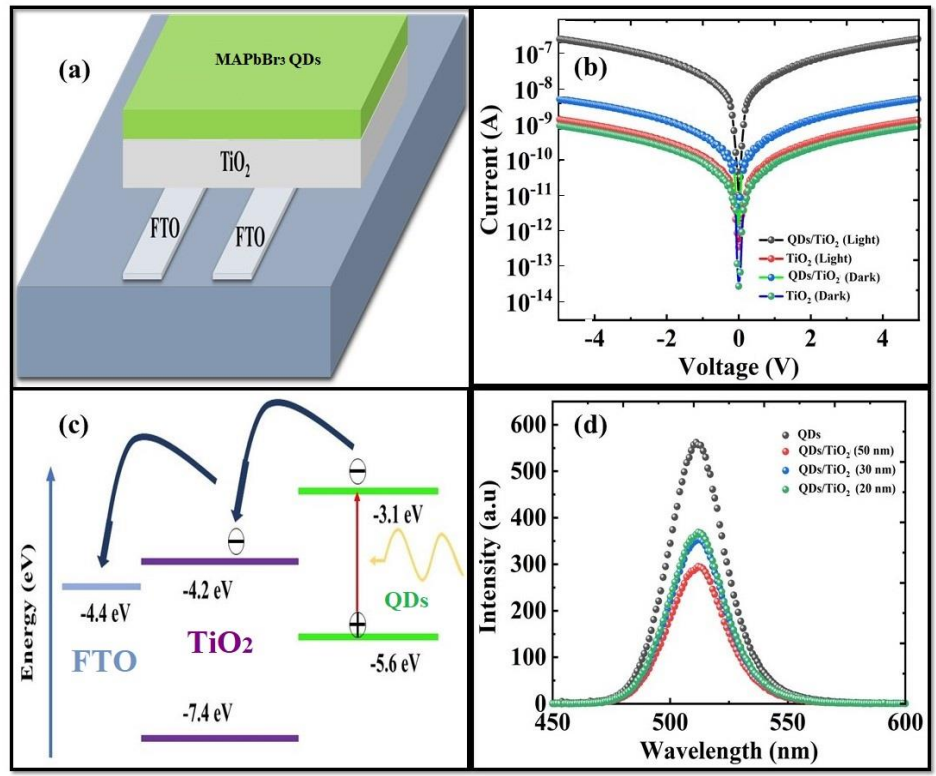

**Figure 2**. (a) Schematic representation of the structure of MAPbBr$_3$ QD/TiO$_2$ heterojunction photodetectors. (b) I-V curves of only TiO$_2$ (PDR) and MAPbBr$_3$ QD/TiO$_2$ heterojunction (PD1) photodetectors (in dark and under light illumination). Current is plotted in log scale. (c) Band diagram of different layers of the heterojunction device demonstrating the electron transfer process under light illumination. (d) PL quenching of perovskite QDs in the presence of TiO$_2$ layer (thicknesses 50, 30 and 20 nm).

## 2.3. MAPbBr$_3$ QD/TiO$_2$ Heterojunction Photodetectors

In addition to PD1, two more devices having thickness of TiO$_2$ layer 30 nm (PD2) and 20 nm (PD3) were fabricated. Similar XRD pattern (Figure 1d) of TiO$_2$ films of varying thickness (50, 30 and 20 nm) infers that there was no crystallographic change of the TiO$_2$ layer with the change of thickness. The photoresponses of all three devices are presented in Figure 3a-c. The device PD1 produces the highest photocurrent ($6.5 \times 10^{-7}$A), while the lowest photocurrent ($4.4 \times 10^{-7}$A) is obtained from PD3. The photocurrents of all the fabricated photodetectors are consistent and reproducible. The on/off ratio of the three devices PD1, PD2 and PD3 are found to be $5.54 \times 10^2$, $5.0 \times 10^2$ and $3.7 \times 10^2$, respectively.

The transient photocurrent behaviour of a photodetector can be understood through the response (rise or decay) time, which is one of the essential performance parameters. The rise time ($T_r$) is defined as the transition time of the photocurrent from 10% to 90% of the peak value, while the decay time ($T_d$) is the time taken to decrease the photocurrent from 90% to 10% of the peak value[14]. Transient photoresponse curves for all three devices are provided in Figure S1 (Supporting Information). The response time of the fabricated devices is calculated and presented in Table 1. Clearly, the response of our perovskite QD photodetector is much faster than recently reported values [15].

Another important photoelectric characteristic of a photodetector is the responsivity (R), which can be defined as electrical output per optical input[16].

$$R = \frac{\Delta I}{PS} \qquad (1)$$

where $\Delta I = I_{on} - I_{off}$, $P$ is the power density of the incident light, and $S$ is the effective area under light exposure. The responsivity is found to be 65, 54 and 45 AW$^{-1}$ for PD1, PD2 and PD3 devices, respectively (Table 1).

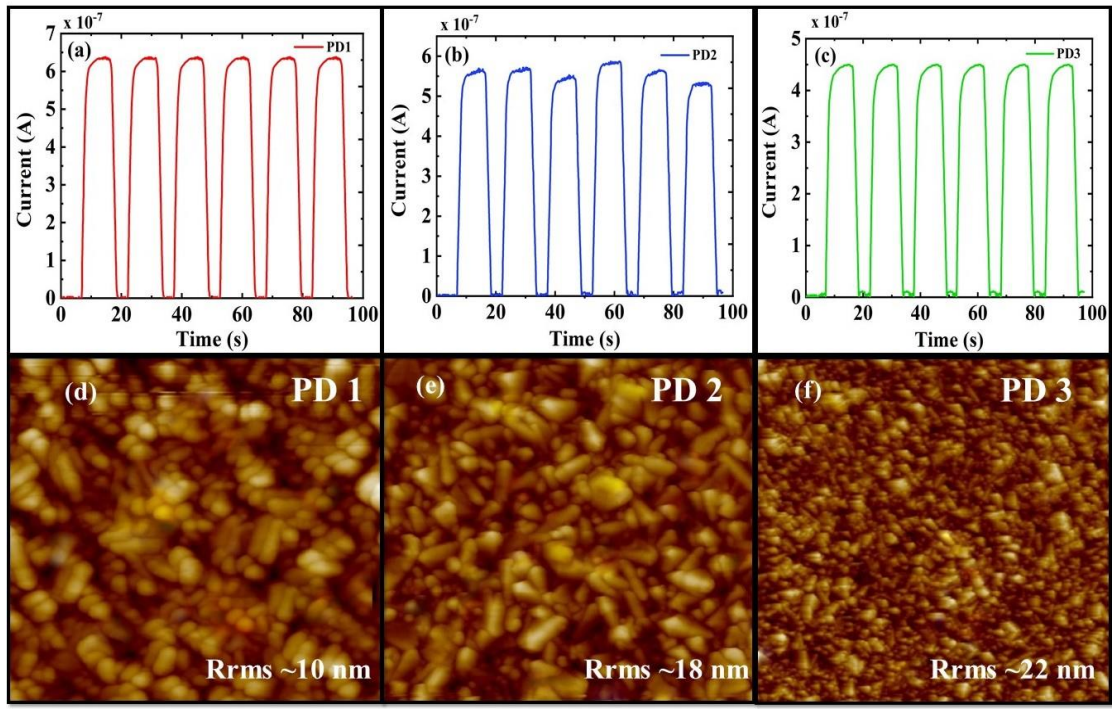

**Figure 3:** (a-c) ON and OFF photoresponsive cycles of three different heterojunction photodetectors. (d-f) AFM images showing surface roughness of TiO$_2$ films of different thicknesses (50, 30 and 20 nm).

It is evident from Table 1 that the overall performance of the bilayer photodetectors depends on thickness of the TiO$_2$ layer and the device PD1 exhibits highest performance. Usually surface roughness of ALD grown TiO$_2$ film is thickness dependent [17]. The AFM images of TiO$_2$ films having thicknesses 50, 30 and 20 nm are shown in Figure 3d-f. The values of surface roughness of these films are 10, 18 and 22 nm, respectively. Clearly, surface roughness varies with the thickness of TiO$_2$ film. Nonetheless, the quenching of PL intensity of MAPbBr$_3$ QDs is more for TiO$_2$ layer of 50 nm thickness (Figure 2d) suggesting much efficient transfer of electrons to the TiO$_2$ film of lowest roughness. The roughness induced surface traps at the interface hinder the electron injection from perovskite to TiO$_2$[17]. Hence, the efficiency of the electron injection to TiO$_2$ layer increases with reduction of roughness of the interface. The improved electron injection increases the electrical conductivity of the TiO$_2$ layer and finally leads to highest photocurrent in the device with thickest TiO$_2$ layer.

**Table 1:** Performance parameters of heterojunction photodetectors (PD1, PD2 and PD3)

| Device | Rise time (s) | Decay time (s) | Responsivity (AW$^{-1}$) | On/Off ratio |
|---|---|---|---|---|
| **PD1** | 0.14 | 0.14 | 65 | 550 |
| **PD2** | 0.20 | 0.15 | 54 | 500 |
| **PD3** | 0.23 | 0.15 | 45 | 370 |

**2.4. Effect of Surface Treatment on the Performance of Heterojunction Photodetectors**

The accumulation of hydrocarbons on the rough surface of the TiO$_2$ layer could deteriorate the quality of the interface affecting the charge transfer across the interface [18]. Several surface treatments such as compact or blocking layer[19], scattering layer[20], atomic doping[21] of the TiO$_2$ films have been testified to enhance solar cell efficiency by improving the interface quality. Treatment with TiCl$_4$ has been applied quite often to increase the efficiency of solar cells and optoelectronic devices [22]. However, TiCl$_4$ is not stable at room temperature as it reacts with moisture present in the air to produce the harmful hydrochloric acid[23]. It is also observed that oxygen ion beam treatment is efficient than oxygen plasma treatment for improving solar cell performance[24]. The main drawback of oxygen plasma treatment is the requirement of considerable capital investment. However, UV-ozone (O$_3$) treatment is an effective and economical surface treatment method [25]. To examine the effect UV-O$_3$ treatment, we fabricated a photodetector with UV-O$_3$ treated TiO$_2$ layer (thickness 50 nm). The I-V characteristics of the device was measured and is presented in Figure S2 (Supporting Information). Time-dependent photocurrent and transient photoresponse curves (measured at 5 V) for ozone untreated and treated photodetectors are presented in figure 4a-c. The treated device exhibits on/off ratio, responsively and rise/decay time of 630, 85 WA$^{-1}$ and 0.09/0.11 s, respectively (Table 2). Undoubtedly, the perovskite QD/TiO$_2$ device displays very high overall performance following the UV-ozone treatment of the TiO$_2$ layer. Recent advancements on the perovskite-based heterojunction photodetectors is summarized in Table S1 (Supporting Information). In comparison to the recent progress, our MAPbBr$_3$ QD/TiO$_2$ heterojunction photodetector has significant advantages over response time and photoresponsivity.

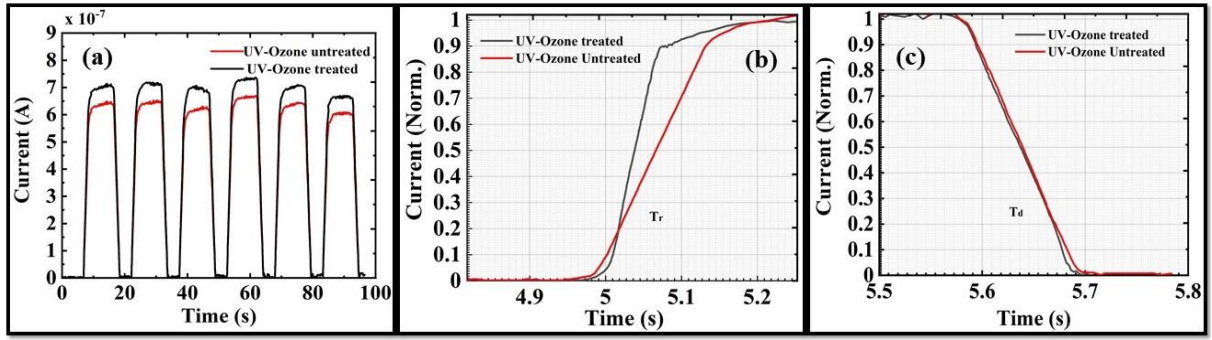

**Figure 4.** (a) Current vs. time graph of UV-ozone treated and untreated photodetectors at 5V. (b-c) Transient photocurrent response (rise/decay time) of ozone-treated and untreated photodetectors.

**Table 2**: Performance parameters of photodetectors before and after ozone treatment

| Device | Rise time (s) | Decay time (s) | Responsivity (AW$^{-1}$) | On/Off ratio |
|---|---|---|---|---|
| **Ozone untreated** | 0.14 | 0.14 | 65 | 550 |
| **Ozone treated** | 0.09 | 0.11 | 85 | 630 |

To understand the effect of UV-ozone treatment, X-ray photoemission spectroscopy (XPS) measurements have been performed for untreated and treated $TiO_2$ films. Figure 5a and b show the C1s peaks of $TiO_2$ films. The intensity of C1s peak at 284.8 eV is found to be reduced after UV-$O_3$ treatment. The quantitative analysis of peaks reveals that the area under C-C peak is decreased after UV-$O_3$ treatment. Clearly, UV-$O_3$ treatment burns out organic contaminants present on the surface of $TiO_2$ film[26] and thereby clean the surface of the film. The O1s peak at 530.1 eV (Figure 5c-d) is associated with oxygen bonding to coordinatively saturated titanium atoms (lattice oxygen of $TiO_2$). The signal at higher binding energy (531.5 eV) could be attributed to the formation of $Ti^{3+}$ surface states through the creation of oxygen vacancies, which is normally inscribed as $Ti_2O_3$[23]. The ratio between the numbers of oxygen atoms present was estimated by comparing areas under $Ti_2O_3$ peaks and found to be increase from 8.2% to 15.4% after the UV-$O_3$ treatment. Therefore, the change of oxidation state from $Ti^{4+}$ to $Ti^{3+}$ takes place through the introduction of negative oxygen species ($O^-$) on the surface of $TiO_2$ during UV-$O_3$ exposure [18b, 26b, 27]. Oxygen vacancies generated during the reduction of $Ti^{4+}$ to $Ti^{3+}$ produces electrons, which affect the surface functionality and charge state of the $TiO_2$ film. The additional electrons generated by UV-$O_3$ treatment improve the carrier transport property of $TiO_2$ film resulting high photocurrent and fast response of the photodetectors. Nonetheless, because of less carbon contaminants, more perovskite QDs could be adsorbed on the $TiO_2$ surface by refining the quality of the heterojunction.

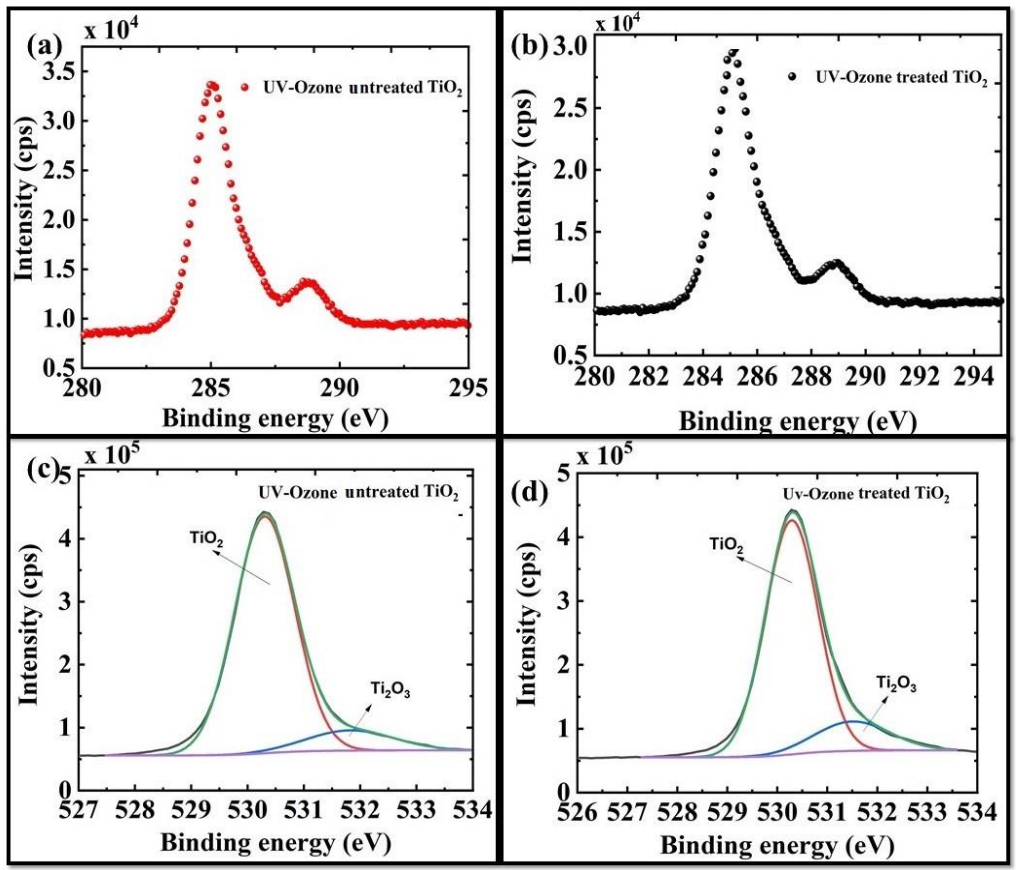

**Figure 5**. XPS spectra of $TiO_2$ films on FTO glass substrates. (a, c) untreated and (b, d) UV-ozone treated $TiO_2$ films. Gaussian/Lorentzian peak fitting technique is used to identify the contribution of individual species.

## 3. Conclusion

In summary, MAPbBr$_3$ QD/TiO$_2$ heterojunction photodetectors were fabricated using atomic layer deposited ultra-thin TiO$_2$ films. Due to the improved conductivity of TiO$_2$ layer through the efficient extraction of electrons at the interface, the performance of heterojunction photodetectors is increased. The performance of such photodetectors depends on the interface quality and the device having lowest interface roughness offers highest performance. UV-ozone treatment of the TiO$_2$ layer enhances the device performance by improving both the interface quality and conductivity of the TiO$_2$ layer. The heterojunction device exhibits high responsivity of 85 AW$^{-1}$, and short rise/decay time of 0.09/0.11 s. In comparison to the reported perovskite-based heterojunction photodetectors, the MAPbBr$_3$ QD/TiO$_2$ device exhibits very high overall performance. Our findings could be beneficial to developing perovskite-based flexible heterojunction photodetectors.

## 4. Experimental Methods

*Materials*: All chemicals were used without further purification. Lead (II) bromide (PbBr$_2$, 99%, Sigma Aldrich), methylamine (CH$_3$NH$_2$, Sigma Aldrich), hydrochloric acid (HCl, 37 wt % in water, Sigma Aldrich), hydrobromic acid (HBr, 48 wt % in water, Sigma Aldrich), oleic acid (Sigma Aldrich), N,N-dimethylformamide (Sigma Aldrich) were used for the preparation of perovskite QDs.

*Synthesis of MAPbBr$_3$ QDs*: Methylammonium bromide (CH$_3$NH$_3$Br) was prepared through a reaction of methylamine in ethanol with HBr at room temperature. HBr was added dropwise while stirring. Upon drying at 100 °C, a white powder of CH$_3$NH$_3$Br was formed. The powder was dried overnight in a vacuum oven and purified with ethanol. To obtain highly green luminescent perovskite QDs, CH$_3$NH$_3$Br and PbBr$_2$ were dissolved in anhydrous dimethylformamide (DMF).

*Device fabrication*: Photodetectors were fabricated on laser patterned fluorine-doped tin oxide (FTO) coated glass substrates (sheet resistivity 7 Ω/cm$^2$) with a channel length of 5 μm. First, substrates were cleaned by ultrasonication in a soap solution, deionized water, acetone, and isopropanol, respectively. The cleaned FTO substrates were baked for 30 minutes before ozone treatment. Then, the ultra-thin layer of TiO$_2$ was deposited by atomic layer deposition (ALD) on FTO at 150 $^0$C. The thickness of the TiO$_2$ films was controlled by varying the deposition cycles. Then perovskite QD films were deposited on the TiO$_2$ layer by spin coating at a speed of 700 rpm for 30 s. The fabricated devices were annealed at 100 $^0$C to enhance the device performance by improving the quality of MAPbBr$_3$QD/TiO$_2$ interface.

*Measurements*: Synthesized QDs were characterized by absorption and photoluminescence (PL) spectroscopy using Shimadzu UV-2450 spectrometer and Cary Eclipse spectrophotometer from Agilent Technologies, respectively. The dimension of the QDs was estimated from transmission electron microscopy (TEM) using TECNAI G2 200 kV (FEI, Electron Optics) electron microscope. X-ray diffraction (XRD) was performed in a Smart Lab, RIGAKU 9 kW rotating anode diffractometer. The surface morphology and thickness of compact TiO$_2$ layer were characterized using atomic force microscopy (Dimension Icon from Bruker) in tapping mode at room temperature and ellipsometry (Accurion EP4), respectively. Current-voltage (I-V) characteristics and photodetection measurements were carried out using a class 3A solar simulator (OAI TriSOL) fitted with an AM (air mass) 1.5 filter for light exposure and equipped with a Keithley 2400 source meter unit. We performed X-ray photoelectron spectroscopy (XPS) in a SPECS instrument with a PHOIBOS 100/150 detector (DLD) at 385 W, 13.85 kV, 39.6 nA (sample current), and a pass energy of 50 eV. Curve synthesis and deconvolution, i.e., identifying the components of the XPS signals, were performed by fitting the XPS signal with Gaussian/Lorentzian function. The binding energies were calibrated using the C1s peak for adventitious carbon at a binding energy of 284.5 eV, with an associated error of ± 0.1− 0.2 eV.

**Supporting Information**

Supporting Information is available from the Wiley Online Library or from the author.

**Acknowledgements**

The experimental facilities in Advanced Materials Research Centre are greatly acknowledged. R.R. and N.N are thankful to IIT Mandi for their fellowship.

# Supporting Information

**High-Performance Perovskite Photodetectors Based on $CH_3NH_3PbBr_3$ Quantum Dot/$TiO_2$ Heterojunction**

**Rajeev Ray, Nagaraju Nakka, and Suman Kalyan Pal[*]**

Advanced Materials Research Centre, India Institute of Technology Mandi, Kamand, Mandi-175075, Himachal Pradesh, India.

School of Basic Sciences, India Institute of Technology Mandi, Kamand, Mandi-1750075, Himachal Pradesh, India

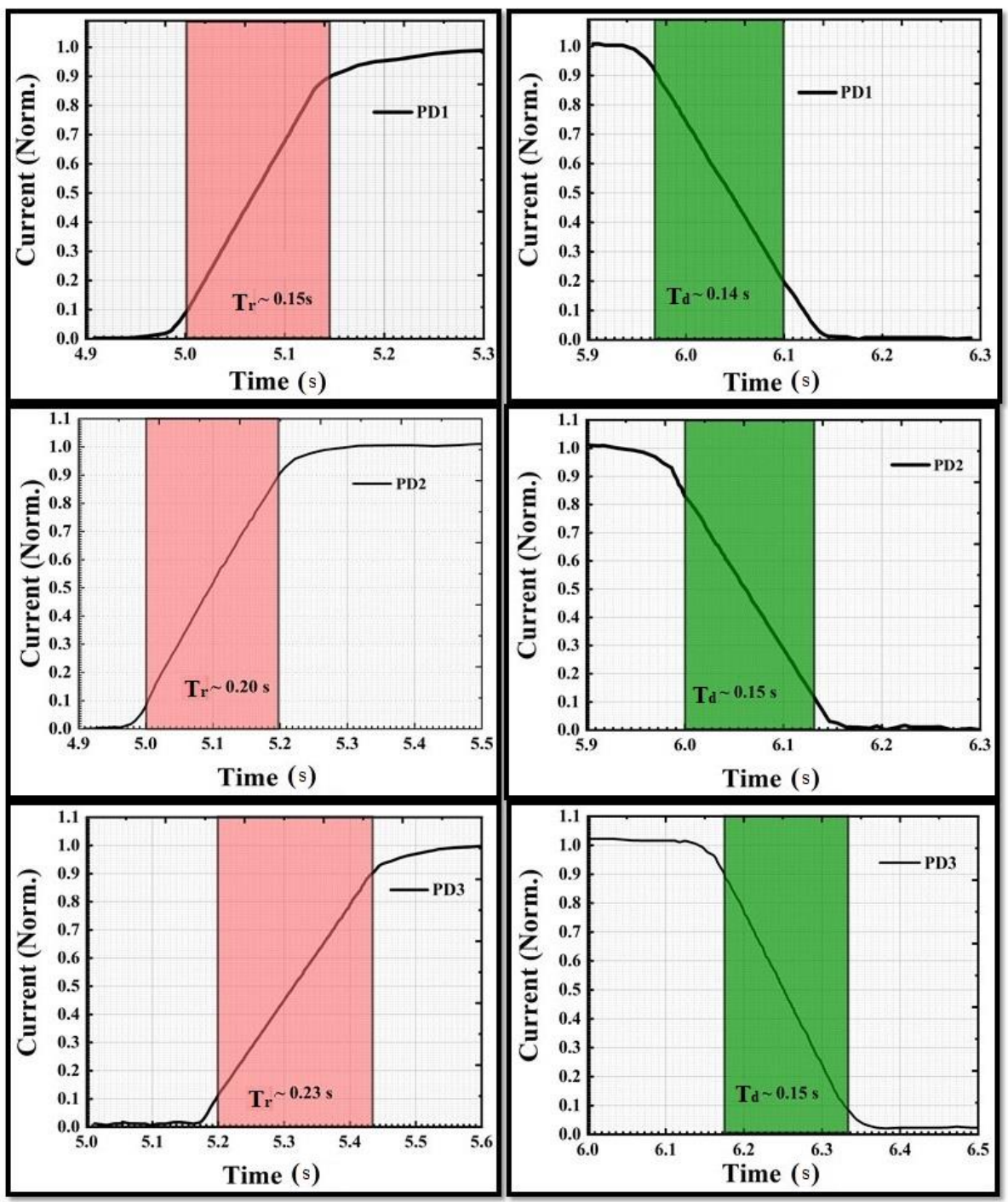

**Figure S1**. Transient photocurrent response of MAPbBr$_3$ QD/TiO$_2$ heterojunction photodetectors (PD1, PD2 and PD3) under white light illumination at a bias of 5 V.

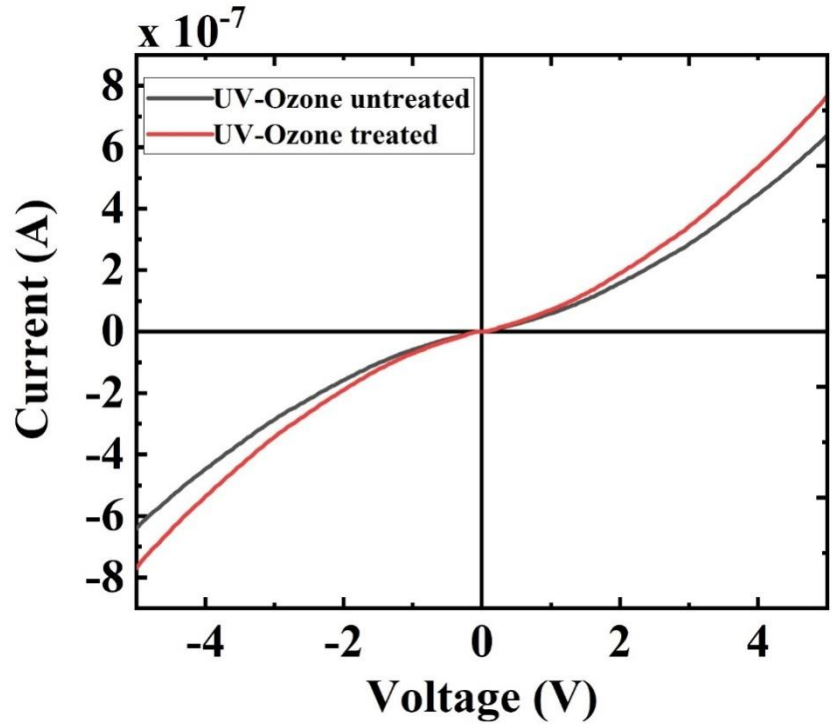

**Figure S2.** I-V characteristics of a photodetector fabricated with UV-ozone untreated and treated TiO$_2$ layer.

**Table S1**. Comparison of response time and responsivity of perovskite-based photodetectors

| Device structure | Rise time (s) | Decay time (s) | Responsivity (AW$^{-1}$) | Ref. |
|---|---|---|---|---|
| TiO$_2$/MAPbBr$_3$ | 0.02 | 0.02 | $0.49 \times 10^{-6}$ | [1] |
| WSe$_2$/MAPbI$_3$ | 2 | 2 | 110 | [2] |
| BiFeO$_3$/CH$_3$NH$_3$PbI$_3$ | 0.74 | 0.09 | 2.0 | [3] |
| ZnO/MAPbI$_3$ | NA | NA | 21.8 | [4] |
| CsPbBr$_3$-MPA (mp-TiO$_2$) | 4.7 | 2.3 | 24.5 | [5] |
| TiO$_2$ Nanocrystal/MAPbBr$_3$ | 0.49 | 0.56 | 0.12 | [6] |
| ZnO NCs/PbS$_x$Se$_{1-x}$ QDs / MAPbBr$_3$ QDs | NA | NA | 16.65 | [7] |
| ALD TiO$_2$/MAPbBr$_3$ QDs | 0.09 | 0.11 | 85.0 | This work |